\newcommand{\ba}{BaCu$_2$Si$_2$O$_7$}

\documentclass[aps,prb,twocolumn,superscriptaddress]{revtex4}
\usepackage{graphicx}

\begin{document}

\title{Structure of the exotic spin-flop states in \ba.}
\author{A. Zheludev}
\email[]{zhelud@bigfoot.com}

\affiliation{Solid State Division, Oak Ridge national Laboratory,
Oak Ridge, TN  37831-6393, USA.} \affiliation{Previous address:
Physics Department, Brookhaven National Laboratory, Upton, NY
11973,USA.}

\author{E. Ressouche}
\affiliation{DRFMC/SPSMS/MDN, CENG, 17 rue des Martyrs, 38054
Grenoble Cedex, France.}

\author{I. Tsukada}
\affiliation{Central Research Institute of Electric Power
Industry, 2-11-1, Iwato kita, Komae-shi, Tokyo 201-8511, Japan}

\author{T. Masuda}
\affiliation{\label{ut}Department of Advanced Materials Science,
The University of Tokyo, 6th Engineering Bldg., 7-3-1 Hongo,
Bunkyo-ku, Tokyo 113-8656, Japan}

\author{K. Uchinokura}
\affiliation{\label{ut}Department of Advanced Materials Science,
The University of Tokyo, 6th Engineering Bldg., 7-3-1 Hongo,
Bunkyo-ku, Tokyo 113-8656, Japan}

\date{\today}

\begin{abstract}
The unusual 2-stage spin flop transition in \ba\ is studied by
single-crystal neutron diffraction. The magnetic structures of the
various spin-flop phases are determined. The results appear to be
inconsistent with the previously proposed theoretical explanation
of the 2-stage transition.
\end{abstract}

\maketitle \narrowtext
\section{Introduction}
The quasi-one-dimensional (quasi-1D) $S=1/2$ antiferromagnet \ba\
is recognized as an almost ideal model material for studying
exotic spin dynamics in weakly-interacting quantum spin
chains.\cite{Tsukada99} In a series of
papers\cite{Zheludev00BaCu2L,Kenzelmann01,BIGC} we have
demonstrated that in this compound it is possible to observe a
separation between single-particle spin wave excitations,
characteristic of a classical system with long-range magnetic
order, and a gapped continuum of states that is a property of the
quantum 1D Heisenberg model. Quite recently however, another
aspect of the physics of \ba\ attracted a great deal of attention.
It was found that in the ordered state of \ba, an application of a
magnetic field along the direction of ordered moment leads to {\it
two} consecutive spin re-orientation transitions.\cite{Tsukada01}
This phenomenon is in contrast with the typical behavior of a
conventional antiferromagnet in a field, where only a single
``spin-flop'' transition occurs.

\begin{figure}
\includegraphics[width=3.2in]{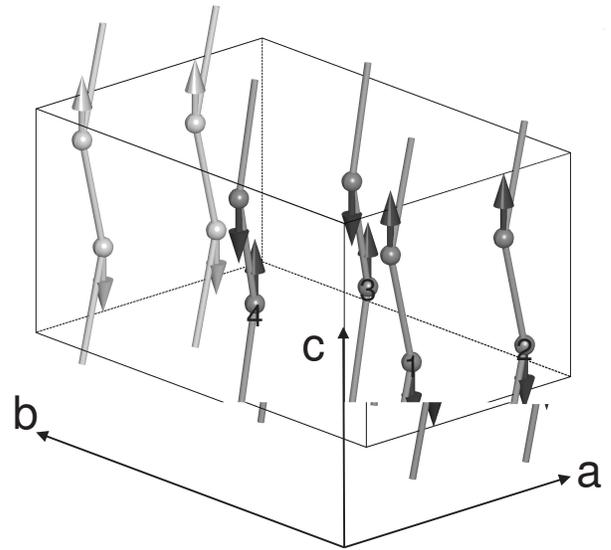}
\caption{\label{struct}A schematic view of the zero-field magnetic
structure of \ba. The $S=1/2$ Cu$^{2+}$ ions form slightly zigzag
antiferromagnetic chains.}
\end{figure}

By far, the largest energy scale of magnetic interactions in \ba\
is that of antiferromagnetic nearest-neighbor coupling within the
spin chains, that run along the $c$ axis of the orthorhombic
crystal structure (space group \textit{Pnma}, $a = 6.862(2)\,$\AA,
$b = 13.178(1)\,$\AA \, and $c =
6.897(1)\,$\AA.\cite{Oliveira93}). The in-chain exchange constant
is $J=24.1$~meV, as previously determined from inelastic neutron
scattering experiments and bulk susceptibility
measurements.\cite{Tsukada99,Kenzelmann01} Exchange coupling
between the chains is much weaker, with $J_x=-0.46$~meV and
$J_y=0.20$~meV, for nearest-neighbors along the $a$ and $b$ axes,
respectively, and $J_3=0.15$~meV along the $(110)$
diagonal.\cite{Kenzelmann01} The spin wave spectrum shows a small
anisotropy gap of about 0.3~meV, corresponding to an easy axis
along the $(001)$ crystallographic direction.\cite{Kenzelmann01}
Long-range magnetic ordering occurs at a low N\'{e}el temperature
of $T_N\approx 9.2$~K.\cite{Tsukada99} The ordered magnetic moment
is rather small, $m_0=$0.15~$\mu_\mathrm{B}$ per
Cu$^{2+}$.\cite{Zheludev00BaCu2L,Kenzelmann01} In the ordered
state the spins are parallel to the magnetic easy axis.
Nearest-neighbor spins along the $b$ and $c$ axes are aligned
antiparallel relative to each other, while nearest neighbors along
$a$ are ferromagnetically aligned.\cite{Tsukada99,Kenzelmann01}
The spin structure is visualized in Fig.~\ref{struct}. The phase
transitions that are the subject of this work occur at relatively
low fields $H_{c1}=2.0$~T and $H_{c2}=4.9$~T, applied along the
easy axis, and produce characteristic jumps in the magnetization
curve.\cite{Tsukada01} To date, the spin structure in the
high-field phases has not been determined. However, in
Ref.~\onlinecite{Tsukada01} it was suggested that upon going
through the first transition, the spins ``flop'' into the $(a,b)$
crystallographic plane. The second transition was then described
as an unusual \textit{spin rotation within the plane perpendicular
to the field}, and is somewhat similar to that recently found in
K$_2$V$_3$O$_8$.\cite{Lumsden01}

Based on an analysis of the high-field magnetization data, it was
conjectured\cite{Tsukada01} that the unusual second spin
re-orientation in \ba\ is a result of competition between
Dzyaloshinskii-Moriya (DM) off-diagonal exchange interaction in
the spin chains and isotropic exchange coupling between the
chains. In the present work we report the results of magnetic
neutron diffraction studies of the various spin-flop states in
\ba. We find the results to be \textit{inconsistent with the
previously proposed model}. It appears that inter-chain coupling
dominates over DM interactions, and that the series of spin
re-orientation transitions can not result from a competition
between the two effects alone. Alternative mechanisms for the
two-stage spin flop transition in \ba\ are discussed.

\section{Experimental}
A $5\times 5 \times 4$~mm$^{3}$ sample was mounted in a 6~T
cryomagnet on the D23 lifting counter diffractometer at Institut
Laue Langevin. The $(0,0,1)$ axis was aligned parallel to the
field direction. With 8 magnetic Cu$^{2+}$ per unit cell, in the
ordered state all magnetic reflections coincide with nuclear ones.
In addition, the ordered magnetic moment in \ba\ is small, and
detecting the magnetic contributions to Bragg intensities is
rather difficult. The measurements were therefore performed in
differential mode. At $H=0$~T, $4$~T (below $H_{c2}$) and 5~T
(above $H_{c2}$) the diffractometer was consecutively positioned
on each accessible Bragg reflection. Without moving any instrument
motors, the peak intensity was measured first at 2~K and then
again at 10~K. This measurement yielded the ratio of nuclear and
magnetic scattering intensities. The absolute values of nuclear
intensities were separately measured at 10~K in standard rocking
scans. The rocking curves also provided an estimate of the
background around each Bragg position. The measured integrated
nuclear intensities, backgrounds, and peak intensity ratios were
then used to estimate integrated magnetic intensities. In another
experiment, the peak intensities of $(\overline{4} 1
\overline{1})$, $(0 \overline{3} \overline{1})$ and $(0
\overline{5} \overline{1})$ reflections were monitored as a
function of applied field in the range 0--5.5~T. Peak intensities
measured in this fashion were assumed to be proportional to
integrated intensities, with the normalization factor determined
at $H=0$, as described above.

 \section{Results}
Due to the strong in-chain coupling, any canting of neighboring
spins away from a perfect antiferromagnetic alignment within each
chain will be small and undetectable in our measurements. In
analyzing the data, we could therefore safely rely on an
approximation in which within each chain all spins are collinear.
Inter-chain coupling is considerably weaker than the in-chain one,
and is comparable in strength to the Zeeman energy associated with
experimental fields. For this reason the relative orientation of
staggered moments in neighboring chains could not be considered
fixed, as it may actually vary with field. Since there are four
chains within each unit cell of \ba, solving the magnetic
structure at each field was reduced to determining the
orientations of four spins, one from each chain, labeled as
$\mathbf{S}_1$--$\mathbf{S}_4$ in Fig.~\ref{struct}.

 \subsection{Spin structure at zero applied field: Phase I}
As a first step, we have verified that our new data collected at
$H=0$ are consistent with the structure previously reported in
Ref.~\onlinecite{Kenzelmann01}. The spins are parallel to the
$c$-axis, with nearest neighbor spins along the $c$ and $b$ axes
antiparallel to each other, and a parallel alignment of nearest
neighbor spins along the $a$ axis, as shown in Fig.~\ref{struct}.
This type of spin arrangement minimizes all inter-chain
interaction energies that were independently measured using
inelastic neutron scattering.\cite{Kenzelmann01} Magnetic Bragg
intensities calculated for this $c$-axis colinear state are in
good agreement with those measured at $H=0$ in the present work.
This comparison is made in Table~\ref{zeroteslat}.
 \begin{table}
 \caption{\label{zeroteslat} Measured
magnetic intensities at $H=0$ in comparison to those calculated
for the $c$-axis colinear state visualized in
Fig.~\protect{\ref{struct}}.}
 \begin{ruledtabular}
 \begin{tabular}{r r r r r r r}
$h$ & $k$& $l$ & $I_{\mathrm{calc}}$ & $I_{\mathrm{obs}}$ &
$\sigma_{\mathrm{obs}}$ &
$\frac{I_{\mathrm{obs}}-I_{\mathrm{calc}}}{\sigma_{\mathrm{obs}}}$\\
\hline
 3 &  -5 &  -1 &      1.073 &      0.014 &      0.349 &     -3.034\\
 -1 &   3 &  -1 &      2.434 &      2.066 &      0.588 &     -0.625\\
 -1 &  -3 &  -1 &      2.547 &      3.574 &      0.605 &      1.697\\
 -1 &  -2 &  -1 &      0.000 &      0.149 &      0.942 &      0.158\\
 -3 &   2 &  -1 &      0.002 &      0.951 &      0.973 &      0.975\\
  4 &   1 &   0 &     19.980 &     19.116 &      1.010 &     -0.855\\
  1 &  -3 &  -1 &      2.547 &      4.038 &      1.015 &      1.469\\
 -4 &   1 &   0 &     19.980 &     20.973 &      1.111 &      0.894\\
 -2 &   6 &  -1 &      0.093 &      0.741 &      1.114 &      0.582\\
 -1 &   2 &  -1 &      0.000 &      2.889 &      1.163 &      2.484\\
  1 &  -1 &  -1 &      1.886 &      2.468 &      1.173 &      0.496\\
  0 &  -6 &   0 &      0.797 &      1.565 &      1.191 &      0.644\\
 -1 &  -1 &  -1 &      1.886 &      3.357 &      0.856 &      1.719\\
 -1 &   1 &  -1 &      1.789 &      4.262 &      1.260 &      1.963\\
  0 &   4 &   0 &      0.527 &      2.109 &      1.352 &      1.170\\
  0 &  -3 &  -1 &     38.129 &     35.631 &      1.080 &     -2.313\\
 -3 &  -2 &  -1 &      0.002 &      1.162 &      1.660 &      0.699\\
 -2 &   1 &   0 &     10.223 &     10.306 &      1.264 &      0.066\\
  0 &  -5 &  -1 &     33.209 &     39.198 &      1.277 &      4.689\\
 -3 &  -5 &  -1 &      1.073 &      3.024 &      1.804 &      1.081\\
 -2 &  -1 &   0 &     10.223 &      9.102 &      1.829 &     -0.613\\
  1 &  -7 &  -1 &      1.421 &      0.974 &      1.341 &     -0.333\\
 -3 &   5 &  -1 &      1.053 &      2.096 &      1.854 &      0.562\\
 -2 &   0 &   0 &      0.000 &      3.193 &      1.897 &      1.683\\
  2 &  -4 &  -1 &      0.058 &      0.306 &      2.009 &      0.123\\
 -2 &   4 &  -1 &      0.057 &      1.714 &      2.106 &      0.787\\
 -1 &  -7 &  -1 &      1.421 &      1.277 &      2.166 &     -0.066\\
 -2 &  -4 &  -1 &      0.058 &      7.373 &      2.227 &      3.284\\
  2 &  -7 &  -1 &     15.898 &     16.154 &      2.626 &      0.098\\
  \hline
  \multicolumn{7}{l}{$\chi^2=2.8$}\\
 \end{tabular}
 \end{ruledtabular}
 \end{table}

 \subsection{Spin structure at $H=4$~T: Phase II}
\begin{figure}
\includegraphics[width=3.2in]{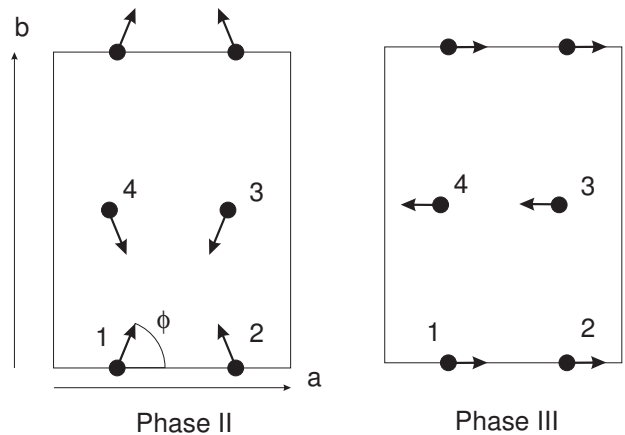}
\caption{\label{fourteslaf}A model for the spin arrangement in
\ba\ in magnetic fields applied along the $c$ axis: Phase II for
$2~\mathrm{T}=H_{c1}<H<H_{c2}=4.7~\mathrm{T}$ and Phase III for
$H>H_{c2}$. The spins are largely confined to the $(a,b)$
crystallographic plane. }
\end{figure}

The magnetic structure at $H=4$~T was solved using a reverse Monte
Carlo analysis, which unambiguously pointed to a planar state with
all spins in the $(a,b)$ plane and a canting by an angle
$\phi\approx80^{\circ}$, as shown in Fig.~\ref{fourteslaf} on the
left. This structure is a slight distortion of a colinear one,
with all spins parallel to the the $b$ axis, and the same relative
alignment of spins in adjacent chains as at $H=0$. A least-squares
refinement of the canting angle and the ordered moment then
yielded $\phi=78(3)^{\circ}$, $m_0=0.18(2)$~$\mu_{\mathrm{B}}$,
with a residual $\chi^2=2.4$.
A comparison between measured and calculated magnetic intensities
is given in Table~\ref{fourteslat}. Assuming $\phi=90^{\circ}$
($b$-axis colinear state), one can still get a reasonably good
agreement with experiment, with $\chi^2=2.8$. Note that the
magnetic structure of Phase II proposed in
Ref.~\onlinecite{Tsukada99} (all spins parallel or antiparallel to
the $a$-axis) is totally inconsistent with the present experiment,
corresponding to $\chi^2=64$, even when $m_0$ is optimized to best
fit the data.
\begin{table}
\caption{\label{fourteslat} Measured magnetic intensities at
$H=4$~T in comparison to those calculated for the canted state
visualized in Fig.~\protect{\ref{fourteslaf}}(left).}
\begin{ruledtabular}
\begin{tabular}{r r r r r r r}
$h$ & $k$& $l$ & $I_{\mathrm{calc}}$ & $I_{\mathrm{obs}}$ &
$\sigma_{\mathrm{obs}}$ &
$\frac{I_{\mathrm{obs}}-I_{\mathrm{calc}}}{\sigma_{\mathrm{obs}}}$\\
\hline
 -4 &   1 &   0 &     14.048 &     13.239 &      0.635 &     -1.273\\
  0 &  -6 &   0 &      0.000 &      0.170 &      1.210 &      0.140\\
 -2 &   0 &   0 &      0.000 &      3.560 &      1.930 &      1.845\\
 -2 &   1 &   0 &      6.126 &     10.760 &      1.750 &      2.648\\
  4 &   1 &   0 &     14.048 &     15.720 &      0.990 &      1.689\\
 -3 &  -5 &  -1 &      0.761 &      1.510 &      1.730 &      0.433\\
 -4 &  -1 &  -1 &     21.136 &     20.890 &      1.340 &     -0.183\\
 -1 &  -7 &  -1 &      0.185 &      2.260 &      2.100 &      0.988\\
 -3 &  -2 &  -1 &      0.447 &      0.710 &      1.620 &      0.163\\
 -1 &  -6 &  -1 &      1.417 &      1.390 &      0.760 &     -0.036\\
 -4 &   1 &  -1 &     21.136 &     22.660 &      1.560 &      0.977\\
 -1 &  -3 &  -1 &      1.487 &      0.730 &      0.800 &     -0.946\\
  0 &  -5 &  -1 &      5.341 &      4.140 &      1.730 &     -0.694\\
 -1 &  -1 &  -1 &      3.828 &      4.330 &      1.180 &      0.426\\
  0 &  -3 &  -1 &     16.846 &     13.950 &      1.480 &     -1.957\\
  2 &  -7 &  -1 &      4.127 &      1.300 &      2.540 &     -1.113\\
 -3 &   5 &  -1 &      0.761 &      2.710 &      1.860 &      1.048\\
  2 &  -6 &  -1 &      0.087 &      0.310 &      0.870 &      0.256\\
 -1 &   1 &  -1 &      3.828 &      8.850 &      1.300 &      3.863\\
  1 &  -3 &  -1 &      1.487 &      0.850 &      1.010 &     -0.631\\
  \hline
  \multicolumn{7}{l}{$\chi^2=2.4$}\\
  \multicolumn{7}{l}{$m_0=0.18(2)$~$\mu_{\mathrm{B}}$}\\
  \multicolumn{7}{l}{$\phi=78(3)^{\circ}$}\\
\end{tabular}
\end{ruledtabular}
\end{table}

\subsection{Spin structure at $H=5$~T: Phase III.}
Fewer data points were collected at $H=5$~T than for the other two
field values (see Table~\ref{fiveteslat}). The best agreement with
the data was obtained assuming a collinear structure with all
spins pointing along the $a$ axis (Fig.~\ref{fourteslaf}, right),
and the same relative alignment of spins in adjacent chains as at
$H=0$. The refined value for the ordered moment is
$m_0=0.17(2)$~$\mu_{\mathrm{B}}$, and for this model $\chi^2=4.5$.
Again we note that the spin arrangement proposed in
Ref.~\onlinecite{Tsukada01} for the high-field phase (all spins
along the $b$ axis and nearest-neighbor spins in the $a$ direction
aligned parallel to each other) is totally inconsistent with
experiment, corresponding to $\chi^2=158$.
\begin{table}
\caption{\label{fiveteslat} Measured magnetic intensities at
$H=5$~T in comparison to those calculated for the $a$-axis
colinear state visualized in Fig.~\protect{\ref{fourteslaf}}
(right).}
\begin{ruledtabular}
\begin{tabular}{r r r r r r r}
$h$ & $k$& $l$ & $I_{\mathrm{calc}}$ & $I_{\mathrm{obs}}$ &
$\sigma_{\mathrm{obs}}$ &
$\frac{I_{\mathrm{obs}}-I_{\mathrm{calc}}}{\sigma_{\mathrm{obs}}}$\\
\hline -4 &   1 &   0 &      0.845 &      0.110 &      0.890 &
-0.826\\
 -4 &  -1 &  -1 &      1.562 &      2.030 &      1.200 &      0.390\\
 -4 &   1 &  -1 &      1.562 &      4.570 &      1.470 &      2.046\\
 -1 &  -3 &  -1 &      3.655 &      2.268 &      0.408 &     -3.396\\
  0 &  -5 &  -1 &     49.258 &     54.301 &      1.672 &      3.016\\
 -1 &  -1 &  -1 &      3.460 &      3.342 &      0.498 &     -0.237\\
  0 &  -3 &  -1 &     68.598 &     61.640 &      2.690 &     -2.587\\
 -1 &   1 &  -1 &      3.460 &      5.706 &      0.876 &      2.562\\
  1 &  -3 &  -1 &      3.655 &      2.200 &      0.725 &     -2.008\\
 -1 &  -6 &  -1 &      0.003 &      0.377 &      0.445 &      0.838\\
   \hline
  \multicolumn{7}{l}{$\chi^2=4.5$}\\
  \multicolumn{7}{l}{$m_0=0.17(2)$~$\mu_{\mathrm{B}}$}\\
\end{tabular}
\end{ruledtabular}
\end{table}

\section{Field dependence.}
\begin{figure}
\includegraphics[width=3.2in]{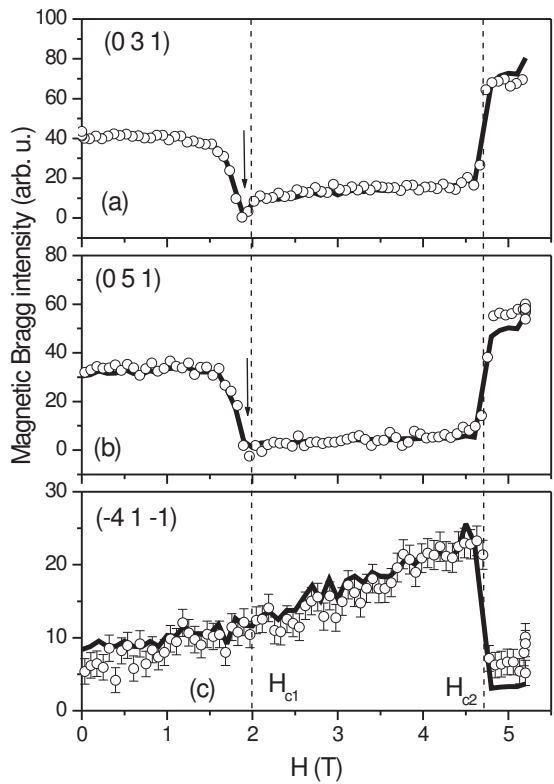}
\caption{\label{peaks} Measured field dependencies of several
magnetic Bragg reflections in \ba\ (symbols). The solid lines are
a fit to the data at each field using the models described in the
text. The arrows indicate intensity dips just below the phase
transition at $H_{c1}$.}
\end{figure}

\begin{figure}
\includegraphics[width=3.2in]{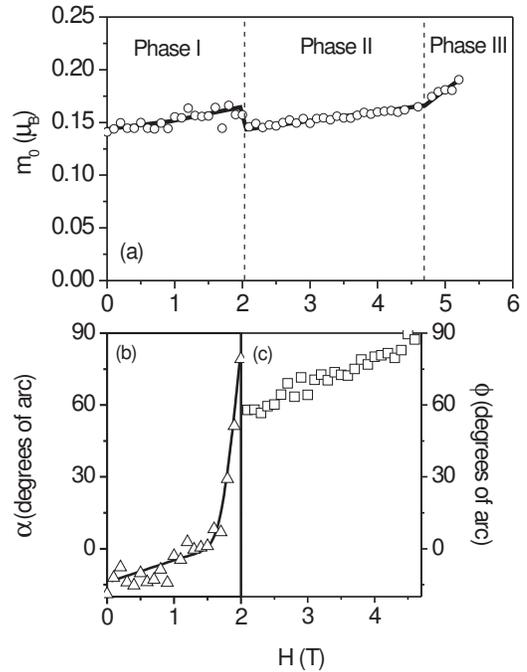}
\caption{\label{params} Symbols: Measured field dependence of
ordered moment (a), tilt angle $\alpha$ in Phase I (b) and canting
angle $\phi$ in Phase II (c). The solid lines are a guide for the
eye.}
\end{figure}

The measured field dependencies of $(\overline{4} 1
\overline{1})$, $(0 \overline{3} \overline{1})$ and $(0
\overline{5} \overline{1})$ reflections are shown in
Fig.~\ref{peaks}. The data for Phases II and III were analyzed
using the established models. The value of magnetic moment $m_0$,
as well as the canting angle $\phi$  for Phase II were refined to
best-fit the three intensities at each field. The results of the
fit are shown in solid lines in Fig.~\ref{peaks}, and the field
dependence of the fit parameters is plotted in Fig.~\ref{params}.

\begin{figure}
\includegraphics[width=2.0in]{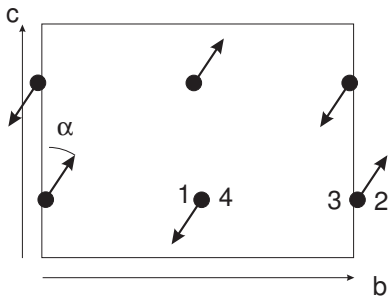}
\caption{\label{tilt} Tilting of the spin structure in \ba\ in
magnetic fields $H<H_{c1}=2$~T, applied along the $c$ axis (Phase
I). The spins are largely confined to the $(b,c)$ crystallographic
plane. }
\end{figure}

To obtain a good fit to the field dependencies measured in Phase
I, an additional parameter had to be introduced to account for the
large change in intensities just below the phase transition at
$H_{c1}$. By trial and error, it was found that the best way to
reproduce this effect is to assume that the spin structure, as a
whole, tilts towards the $b$ axis (Fig.~\ref{tilt}). In this
model, when the direction of staggered moment passes through the
$(0 \overline{3} \overline{1})$ direction, the corresponding Bragg
intensity shows a characteristic dip (arrow in Fig.~\ref{peaks}a).
This behavior is a result of the neutron polarization intensity
factor going to zero for scattering vectors that are parallel to
ordered magnetic moment. A smaller dip, at a slightly larger field
is seen in the $(0 \overline{5} \overline{1})$ peak as well (arrow
in Fig.~\ref{peaks}b). By varying $m_0$ and the tilt angle
$\alpha$, a good fit to the data collected at $0<H<H_{c1}$ could
be obtained. The result of the fit is shown in solid lines in
Fig.~\ref{peaks}, and the field dependence of the fit parameters
is plotted in Fig.~\ref{params}.
Note that the angle $\alpha$ is shown as negative at small fields.
In fact, the tilt $\alpha$ probably remains close to zero and
positive, except immediately before the transition, where it
rapidly increases to $90^{\circ}$. The negative values obtained in
the fit are an artifact of using only three measured intensities
in the analysis. For small $\alpha$ the experiment is fairly
insensitive to this parameter, due to a $\cos^2\alpha$ term in the
intrinsic polarization factor for the scattering cross section and
the fact that all data were collected close to the $(001)$
reciprocal-space plane. However, as $\alpha$ approaches
$90^{\circ}$, the polarization sensitivity is greatly enhanced.

\section{Discussion}
The data analysis described above was based on the known
room-temperature crystal structure. The use of room-temperature
structure factors to determine spin orientations at low
temperatures may lead to systematic errors. Additional errors may
be associated with measuring magnetic intensities in the
temperature-differential mode. Nevertheless, the results allow us
to establish the main features of the three phases: 1)
Nearest-neighbor spins within each chain remain almost
anti-parallel to each other at all times. 2) The relative
alignment of adjacent spins from neighboring chains does not
deviate significantly from that at $H=0$, with ferromagnetic spin
alignment along $a$ and antiferromagnetic along $b$. Even in Phase
II the canting is not very dramatic and vanishes just before the
transition at $H_{c2}$. 3) The spins are oriented roughly along
the $c$, $b$ and $a$ axis in phases I, II and III, respectively.
4) The ordered moment is only slightly field-dependent and remains
close to $0.15\mu_{\mathrm{B}}$.

In weakly coupled quantum spin chains the ordered moment $m_0$ is
proportional to $\sqrt{|J_{\bot}|/J}$, where $J_{\bot}$ is the
effective mean-field inter-chain coupling strength.\cite{Schulz96}
At $H=0$ the structure is such that
$|J_{\bot}|=J_b+2J_3+|J_a|$.\cite{Kenzelmann01} The fact that
$m_0$ remains practically constant when going through the phase
transitions, means that in Phases II and III, $J_{\bot}$ does not
change much and nearest-neighbor spins remain close to being
parallel to one another along the $a$ axis and antiparallel along
the $b$ and $c$ directions. This can be seen as an important
self-consistency check for our determination of the spin
orientations in the three phases. The overall gradual increase of
the ordered moment as a function of $H$  seen in
Fig.~\ref{params}a is to be expected, and is due to the
suppression of 1D quantum spin fluctuations by the
symmetry-breaking external field. The slight suppression of $m_0$
observed in Phase II (Fig.~\ref{params}a) is a result of the
canted geometry, in which $|J_{\bot}|$ is reduced compared to that
in Phase I. In the presence of canting ($\phi<90^{\circ}$) the
effective inter-chain coupling is given by
$|J_{\bot}|=2J_3-(J_b+|J_a|)\cos(2\phi)$. In particular, just
above $H_{c1}$, $\phi= 60^{\circ}$, and the drop in $m_0$ expected
on going through the first phase transition due to the abrupt
change in $|J_{\bot}|$ is about 23\%. This value is in reasonable
agreement with the observed $\approx 15\%$ effect.

Given the available data, the spin re-orientation at $H_{c1}$ can
be described as a more or less conventional spin-flop transition,
that one expects for a classical antiferromagnet in a magnetic
field applied along the easy axis. In zero field, the
easy-$c$-axis anisotropy energy is minimized when all spins are
perpendicular to the $(a,b)$ plane. To take advantage of the
Zeeman energy in applied fields, the spins have to align
themselves perpendicular to $c$, to allow a canting of the rigid
antiferromagnetic structure in the direction of $\mathbf{H}$.
Dzyaloshinskii-Moriya interactions and inter-chain coupling may
modify the transition field and produce the small canting found in
Phase II, but do not affect the basic physics of the effect. The
fact that the transition appears to have a precursor, the spin
structure tilting by an angle $\alpha$ at $H<H_{c1}$, may have
several explanations. One possibility is that this phenomenon is
result of an imperfect alignment of the external magnetic field
relative to the $c$ axis of the crystal. Alternatively, it can be
an intrinsic effect, due to the presence of Dzyaloshinskii-Moriya
interaction, or to the existing structural canting of the local
anisotropy axes of Cu$^{2+}$ relative to the $c$ axis. To fully
resolve this problem, a careful determination of the magnetic
structure just below and just above $H_{c1}$ will be required.

The second phase transition is much more unusual, since it
involves a spin rotation in the plane perpendicular to
$\mathbf{H}$. It is clear that such behavior requires the presence
of off-diagonal exchange interactions in the system. As recently
observed in K$_2$V$_3$O$_8$, for example, a spin rotation around
the field direction can be caused by a competition between DM
interactions and magnetic anisotropy.\cite{Lumsden01} This type of
behavior is known since earlier studies of
hematite.\cite{hematite} For \ba\ though, a totally different
explanation was proposed in Ref.~\onlinecite{Tsukada01}, and the
transition was attributed to a competition between DM coupling in
the chains and isotropic inter-chain interactions, a mechanism
similar to that previously discussed in relevance to the anomalous
spin-flop behavior of La$_2$CuO$_4$.\cite{Thio90}

The key argument of Ref.~\onlinecite{Tsukada01} is that the main
components of the Dzyaloshinskii vector for the nearest-neighbor
bond within the chains lie in  the $(a,b)$ plane, and alternate
sign from one bond to the next. In this case, DM interactions
produce a weak-ferromagnetic canting of the spins within each
chain even in zero applied field. In the presence of an external
field, additional canting is due to Zeeman energy. The free energy
of the system is minimized when the two canting effects are in the
same direction. For this to be the case, nearest-neighbor spins
along the $a$ axis have to be almost \textit{antiparallel} to each
other, since Dzyaloshinskii vectors in the adjacent chains are
\textit{antiparallel} as well. This preferred antiparallel
orientation is in competition with \textit{ferromagnetic}
inter-chain coupling along the $a$ axis. In the model proposed in
Ref.~\onlinecite{Tsukada01}, the frustration is resolved in Phase
II by having all spins aligned almost parallel to the
Dzyaloshinskii vector (roughly along the $a$ axis), which
practically eliminates the Dzyaloshinskii energy altogether, and
the ferromagnetic inter-chain coupling energy along the $a$ axis
is minimized. In Phase III the model predicts that the combination
of DM and Zeeman energies wins over inter-chain interactions: the
spins rotate to be almost parallel to the $b$ axis and cant in the
direction of the applied field to minimize both energies. At the
same time, nearest-neighbors along the $a$ axis become almost
antiparallel, despite a ferromagnetic coupling in this direction.

The results presented above clearly demonstrate that the actual
spin arrangements in \ba\ in Phases II and III are totally
different from those previously conjectured. In
Ref.~\onlinecite{Tsukada01} it was emphasized that the
Dzyaloshinskii vectors not being strictly parallel to the $a$
axis, and additional inter-chain interactions along the $(0,1,0)$
and $(1,1,0)$ directions being present, the critical field and the
details of the spin structures may deviate from those predicted by
the simplified model. However, even in the more general case, if
the $H_{c2}$ transition in \ba\ was indeed driven by a competition
between DM interactions and isotropic inter-chain coupling, the
energy of the latter would have to change abruptly upon going
through the transition. The present data clearly show that this is
{\it not} the case, since at $H_{c2}$ the spin structure rotates
as a whole, and the relative spin orientations, and with them the
inter-chain exchange energies, remain practically unchanged. It
appears that inter-chain coupling dominates over DM interactions
in all three phases. Because of the canting in Phase II, the
energy of  inter-chain interactions is actually better minimized
in Phase III, rather than Phase II. This sequence is opposite to
that emerging from the model proposed in
Ref.~\onlinecite{Tsukada01}.

We believe that at the present stage it is unwise to attempt a
quantitative explanation of the phase transitions in \ba\ based on
some specific spin Hamiltonian. Too many parameters are involved
to make such an analysis unambiguous. The energy scale of
inter-chain interactions and $c$-axis anisotropy are indeed
similar to that defined by transition fields. However, the
Hamiltonian is expected to have other terms on the same energy
scale, and these {\it can not be dismissed}. In particular,
measurements of the spin wave spectrum contain evidence of an
in-plane anisotropy of the magnitude $\approx0.15$~meV. This
anisotropy is most likely due to the so-called Kaplan Schekhtman
Entin-Wohlman Aharony (KSEA) interactions.\cite{KSEA} The KSEA
term is a companion to DM interactions, and is an easy axis
parallel to the Dzyaloshinskii vector. KSEA interactions are
believed to be the driving force in the spin reorientation
transition in K$_2$V$_3$O$_8$.\cite{Lumsden01} The similarity of
the behavior of the latter compound with that found in \ba\ may
indicate that anisotropy effects, and KSEA interactions in
particular, could be responsible for the $H_{c2}$ transition in
the silicate as well.

\section{Conclusion}
In summary, the actual mechanism of the exotic two-stage spin flop
transition in \ba\ remains a mystery. What is clear though, is
that DM interactions and inter-chain exchange coupling alone can
not sufficient to explain phenomenon. It appears more likely that
both transitions in \ba\ are caused by a competition between DM
interactions and diagonal magnetic anisotropy effects, of which
more needs to be learned.

\begin{acknowledgments}
One of the authors (I. T.) thanks J. Takeya for useful
discussions. This work is supported in part by the Grant-in-Aid
for COE Research ``SCP coupled system" of the Japanes Ministry of
Education, Science, Sports, and Culture. Oak Ridge National
Laboratory is managed by UT-Battelle, LLC for the U.S. Department
of Energy under contract DE-AC05-00OR22725.
\end{acknowledgments}

\end{document}